# Enrichment by Extragalactic First Stars in the Large Magellanic Cloud

**Anirudh Chiti[1,2], Mohammad Mardini[3,4,5,6], Guilherme Limberg[7,1,2], Anna Frebel[5,6], Alexander P. Ji[1,2], Henrique Reggiani[8], Peter Ferguson[9], Hillary Diane Andales[5,6], Kaley Brauer[5,10], Ting S. Li[11,12], Joshua D. Simon[8]**

The Large Magellanic Cloud (LMC) is the Milky Way's most massive satellite galaxy[1,2], which only recently (~2 billion years ago) fell into our Galaxy[3]. Since stellar atmospheres preserve their natal cloud's composition[4,5], the LMC's recent infall makes its most ancient, metal-deficient ("low-metallicity") stars unique windows into early star formation and nucleosynthesis in a formerly distant region of the high-redshift universe. Previously, identifying such stars in the LMC was challenging[6]. But new techniques have opened this window[7,8,9], now enabling tests of whether the earliest element enrichment and star formation in distant, extragalactic proto-galaxies deviated from what occurred in the proto-Milky Way. Here we present the elemental abundances of 10 stars in the LMC with iron-to-hydrogen ratios ranging from ~1/300th to ~1/12,000th of the Sun. Our most metal-deficient star is 50 times more metal-deficient than any in the LMC with available detailed chemical abundance patterns[6], and is likely enriched by a single extragalactic first star

[1] Department of Astronomy & Astrophysics, University of Chicago, 5640 S. Ellis Avenue, Chicago, IL 60637
[2] Kavli Institute for Cosmological Physics, University of Chicago, Chicago, IL 60637, USA
[3] Department of Physics, Zarqa University, Zarqa 13110, Jordan
[4] Institute for AI and Beyond, The University of Tokyo, 7-3-1 Hongo, Bunkyo-ku, Tokyo 113-8655, Japan
[5] Joint Institute for Nuclear Astrophysics—Center for the Evolution of the Elements (JINA-CEE), USA
[6] Department of Physics and Kavli Institute for Astrophysics and Space Research, Massachusetts Institute of Technology, Cambridge, MA 02139, USA
[7] Universidade de São Paulo, Instituto de Astronomia, Geofísica e Ciências Atmosféricas, Departamento de Astronomia, SP 05508-090, São Paulo, Brasil
[8] The Observatories of the Carnegie Institution for Science, 813 Santa Barbara Street, Pasadena, CA 91101, USA
[9] Department of Physics, University of Wisconsin-Madison, Madison, WI 53706, USA
[10] Center for Astrophysics | Harvard & Smithsonian, Cambridge, MA 02138, USA
[11] Department of Astronomy and Astrophysics, University of Toronto, 50 St. George Street, Toronto ON, M5S 3H4, Canada
[12] Dunlap Institute for Astronomy & Astrophysics, University of Toronto, 50 St George Street, Toronto, ON M5S 3H4, Canada

**supernova[10,11]. This star lacks significant carbon-enhancement, as does our overall sample, in contrast with the lowest metallicity Milky Way stars[12]. This, and other abundance differences[13], affirm that the extragalactic early LMC experienced diverging enrichment processes compared to the early Milky Way. Early element production, driven by the earliest stars, thus appears to proceed in an environment-dependent manner.**

We performed a search for low-metallicity stars in the LMC by applying metallicity-sensitive photometric selections to data from Data Release 3 (DR3) of the Gaia mission[14]. We selected likely LMC member stars by identifying all stars within 10° of its center that had Gaia DR3 proper motions consistent with its bulk motion[13,15]. We then derived metallicities for these stars by calculating their simulated flux in metallicity-sensitive passbands[8,16] using spectrophotometric data from Gaia BP/RP spectra[17]. These fluxes were compared to fluxes from synthetic stellar spectra at various metallicities[9] (see methods) to select stars with metallicities below that of the previously known most metal-poor red-giant star in the LMC with detailed chemical abundances ([Fe/H] = -2.4[6], where [Fe/H] is defined as the logarithmic iron-to-hydrogen ratio relative to the Sun). Figure 1 outlines this selection procedure, starting with the spatial distribution of these stars (panel a), the color-magnitude diagram of the LMC field (b), proper motions of these stars (c), and a color-color plot to flag low-metallicity candidates (d).

We obtained follow-up spectra of these candidates using the high-resolution Magellan Inamori Kyocera Echelle (MIKE) spectrograph[18] on the 6.5m Magellan-Clay telescope. We identified ten stars with metallicities extending from -4.2 < [Fe/H] < -2.5, nearly two orders of magnitude below LMC stars with previously determined detailed elemental abundance measurements[6,13] and significantly more metal-poor than any previously detected[7]. We derived abundances for up to 23 elements for our sample using standard one-dimensional (1D) stellar model atmospheres under the assumption of local thermodynamic equilibrium (LTE)[19]. The radial velocities measured from these spectra confirmed that these stars are gravitationally bound members of the LMC, shown by integrating their orbits in a gravitational potential model that included both the LMC and the Milky Way[20] (see methods).

Star LMC-119 has the lowest metallicity in our sample ([Fe/H] = -4.13±0.20) and is the lowest metallicity star known in any external galaxy[21]. Given its extremely low metallicity, this star exhibits the characteristics of a second-generation star that preserves the chemical imprints of a first star supernova[10,11]. Highlighting its uniqueness, the LMC's earliest stars likely formed when it was at a comoving distance of ~1 Mpc to ~3 Mpc away from the Milky Way (see methods), at the distance scale of large scale structure variations in the early universe[22]. Star LMC-119 thus provides a rich opportunity to explore to what extent local first stars differ in their nature across different portions of early large scale structure through comparisons to analogous Milky Way stars. We fit the chemical abundances of LMC-119 using Population III supernova yield models from ref. [23] and find that they are reproduced by supernovae with progenitor masses between 10 to 50 times the mass of the Sun and explosion energies up to $4 \times 10^{51}$ ergs (see methods). The chemical composition of LMC-119 broadly aligns with what is seen in second-generation Milky Way halo stars (see panel d in Figure 2), except that we find an unusually low carbon abundance for this star ([C/Fe] < 0.3). For reference, ~90% of Milky Way stars with [Fe/H] < -4.0 have [C/Fe] > +0.7 (known as "carbon-enhanced metal-poor" stars)[24]. This discrepancy hints at a possible divergence in the production channels of carbon, whose overwhelming overabundance in Milky Way halo stars is often taken as a characteristic signature of the first stars[25]. But there have been recent glimpses that carbon-enhancement may not be a universal outcome of early element enrichment: one other [Fe/H] ~ -4.0 star located in a dwarf galaxy[21] is not carbon-enhanced, nor are stars in the Galactic bulge with higher [Fe/H] [26,27], as well as a handful of the most metal-deficient Milky Way halo and disk stars[24,28,29,30].

Strikingly, we find that none of our LMC low-metallicity stars are carbon-enhanced metal-poor stars (see Figure 2, panel b). This is in significant tension with the Milky Way halo where carbon-enhancement is frequent (25% when [Fe/H] < -2.5)[12]. There is a

6% probability that an unbiased sample of 10 stars below [Fe/H] = -2.5 would show no carbon enhancement. To further investigate this discrepancy, we observed an additional 15 stars in the LMC, of which 8 had [Fe/H] < -2.5, with short exposures to solely derive [Fe/H] and [C/Fe]. None of these stars are carbon-enhanced either (see methods). Combining these samples leads to a <1% probability of finding no stars that are carbon enhanced below [Fe/H] = -2.5. While the selection may have missed the most C-enhanced stars in the LMC with [C/Fe] > 1.3 (see methods), this analysis suggests that the statistical frequency of stars with [C/Fe] > 0.7 through [C/Fe] ~ 1.3 is lower than the Milky Way. The lack of this carbon-enhanced tail of the distribution suggests that some sites of high carbon production (e.g., faint supernovae[31] and winds of fast-rotating massive stars[32]) in the early Milky Way ecosystem may not have been nearly as dominant in the LMC.

Other distinct LMC chemical abundance ratios demonstrate additional differences from equivalent ratios found in the Milky Way and its satellite galaxies. First, the LMC appears to have had strikingly inefficient early star formation or inflows of gas, affirming evidence from its more metal-rich stars[6,13] from the abundances of its α-elements (e.g., Mg, Ca) as a function of Fe abundance. The downturn in this trend traces the efficiency of star formation (see methods) and including our LMC sample with the literature, we model this downturn with a simple linear model and find that the initial decline occurs at [Fe/H] < -1.82. This is significantly lower than the location of the downturn in the Milky Way ([Fe/H] > -1.0[33]) and what is expected given the LMC's mass ([Fe/H] ~ -1.2[13]). Interestingly, this is comparable to other Milky Way satellites that are three to five orders of magnitude less massive[34,35] than the LMC (see top panel in Figure 2). Second, prior work in the LMC noted that stars between -2.5 < [Fe/H] < -1.5 were over-enhanced in europium and other elements produced by the rapid (r-) neutron-capture process[6]. We find no evidence for europium enhancement ([Eu/Fe] > 0.7) in any of our low metallicity stars (see Figure 2, panel c), in contrast to the LMC sample with [Fe/H] > -2.5. A likely explanation is that r-process events occurred in the LMC but only at a delayed time once its gas had been enriched to [Fe/H] ~ -2.5. This is consistent with inefficient star formation, which may lessen the likelihood that rare sites of r-process nucleosynthesis are produced at the lowest metallicities due to slower stellar mass production[6]. The collectively observed differences thus point to significant diversity of early chemical and galactic evolution scenarios. After all, the LMC inhabited a qualitatively distant local environment in its early evolution.

We find intriguing glimpses that a subset of our stars may have been accreted into the LMC from smaller progenitor systems. This confirms prior suggestions of mass assembly using globular cluster data[36]. We computed the proper motion of our stars relative to the reference frame of the LMC on the plane of the sky, and found that 8/10 are rotating directionally with the bulk motion of the galaxy, but 2/10 are counter-rotating ( extended data Figure 2). One way to form counter-rotating stars is through accretion events or a collision with the Small Magellanic Cloud[37,38], and future work may use the chemistry of lowest metallicity stars in the Magellanic Cloud ecosystem as a whole to investigate processes arising from past encounters between the Large and Small Magellanic Clouds. Additionally, a handful of our stars also show a deficiency in Barium and Strontium ([Ba/Fe] $\lesssim$ -1.0, [Sr/Fe] $\lesssim$ -1.0), which is an elemental signature typically seen in stars that formed in smaller, ultra-faint dwarf galaxies[39]. Regardless, these stars probe gas from the early ecosystem of the LMC, as preserved by the chemically pristine stars in its own stellar halo. This showcases the powerful insights to be gained from extending the galactic stellar archaeology framework to the LMC to unveil the assembly of its halo. Given the high success rate in finding lowest-metallicity stars in the LMC (see methods), our largest satellite galaxy is now a new frontier for stellar archaeology, extending this field to extragalactic scales.

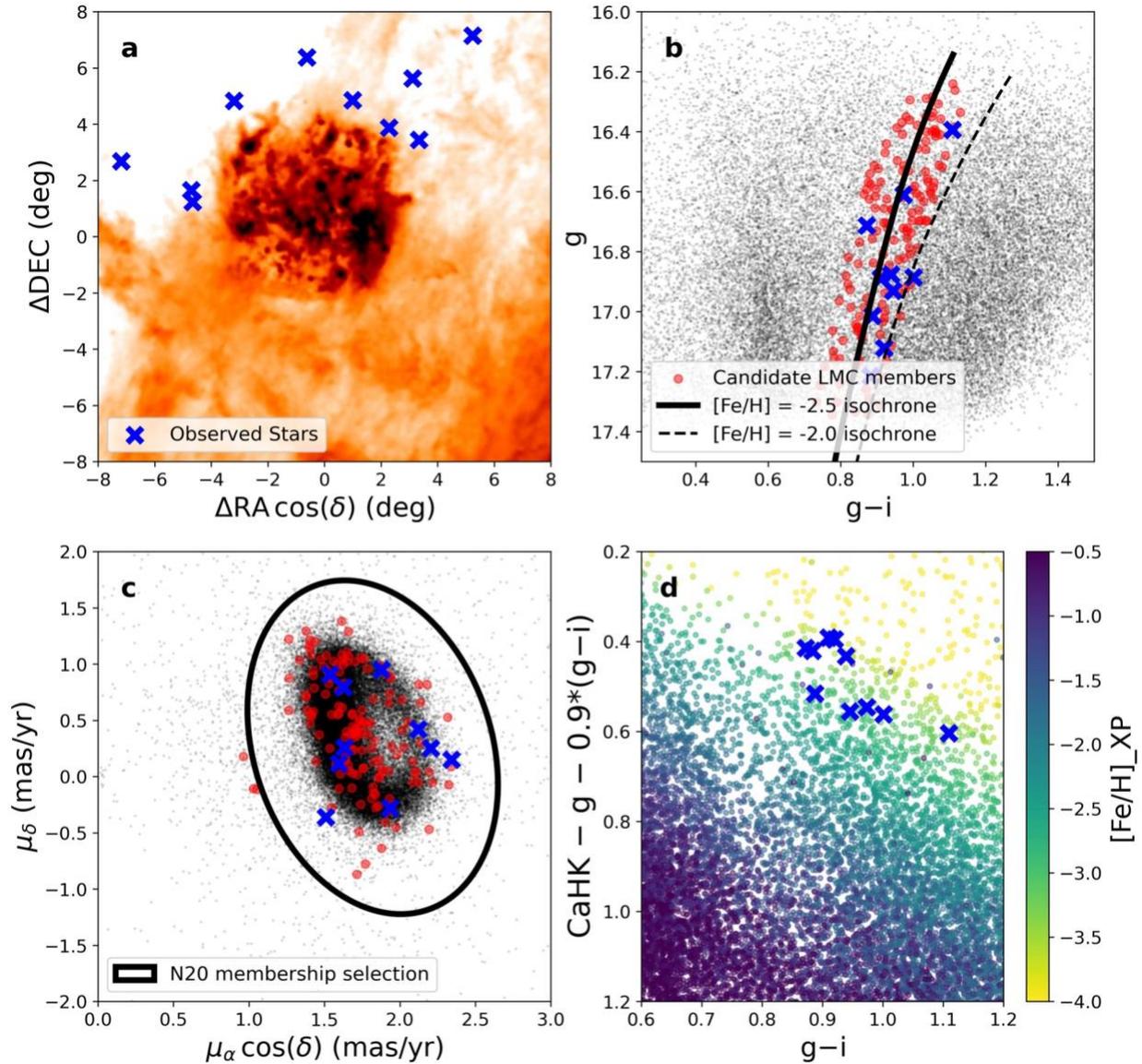

**Figure 1:** Identification of low-metallicity member stars in the Large Magellanic Cloud (LMC).

**a.** Locations of low-metallicity LMC member stars observed in this study are indicated by blue crosses. Coordinates are centered on the LMC (RA = 5h23m34s, DEC = -69d45m22s). The colors in the background indicate reddening E(B-V) values from a canonical dustmap[40]. We avoid regions of high reddening in our target selection, since these regions suppress the signal in the blue region of the Gaia BP/RP spectra.
**b.** A color-magnitude diagram of stars within 10° of the LMC, with tracks ("isochrones") corresponding to a 12 Gyr, [Fe/H] = -2.5 stellar population (solid black line) and 12 Gyr, [Fe/H] = -2.0 stellar population (dashed black line)[41]. All stars are shown as grey points, stars with LMC-consistent proper motions and locations consistent with the [Fe/H] = -2.5 isochrone are shown in red. Our observed stars are, again, shown as blue crosses.

**c.** Proper motions of stars within 10° of the LMC center, with the boundary of the membership selection in ref. [13] denoted as a black ellipse. As in prior panels, all stars are shown in gray, those passing a proper motion and isochrone selection are shown in red, and observed stars are blue crosses.

**d.** Metallicity-sensitive color-color plot of all stars within 10° of the LMC, generated from synthetic photometry from the Gaia XP spectra using XPytools[14,17]. Each datapoint is colored by the metallicity as inferred from this color-color plot using methods in ref. [9]. Note that all stars that were observed, as indicated by the blue crosses, occupy the low metallicity space in this color-color plot. For clarity, we also note that avoiding stars that e.g., have large flux uncertainties, are faint, or have nearby companions shaped our choice of candidates to observe from the low metallicity candidates.

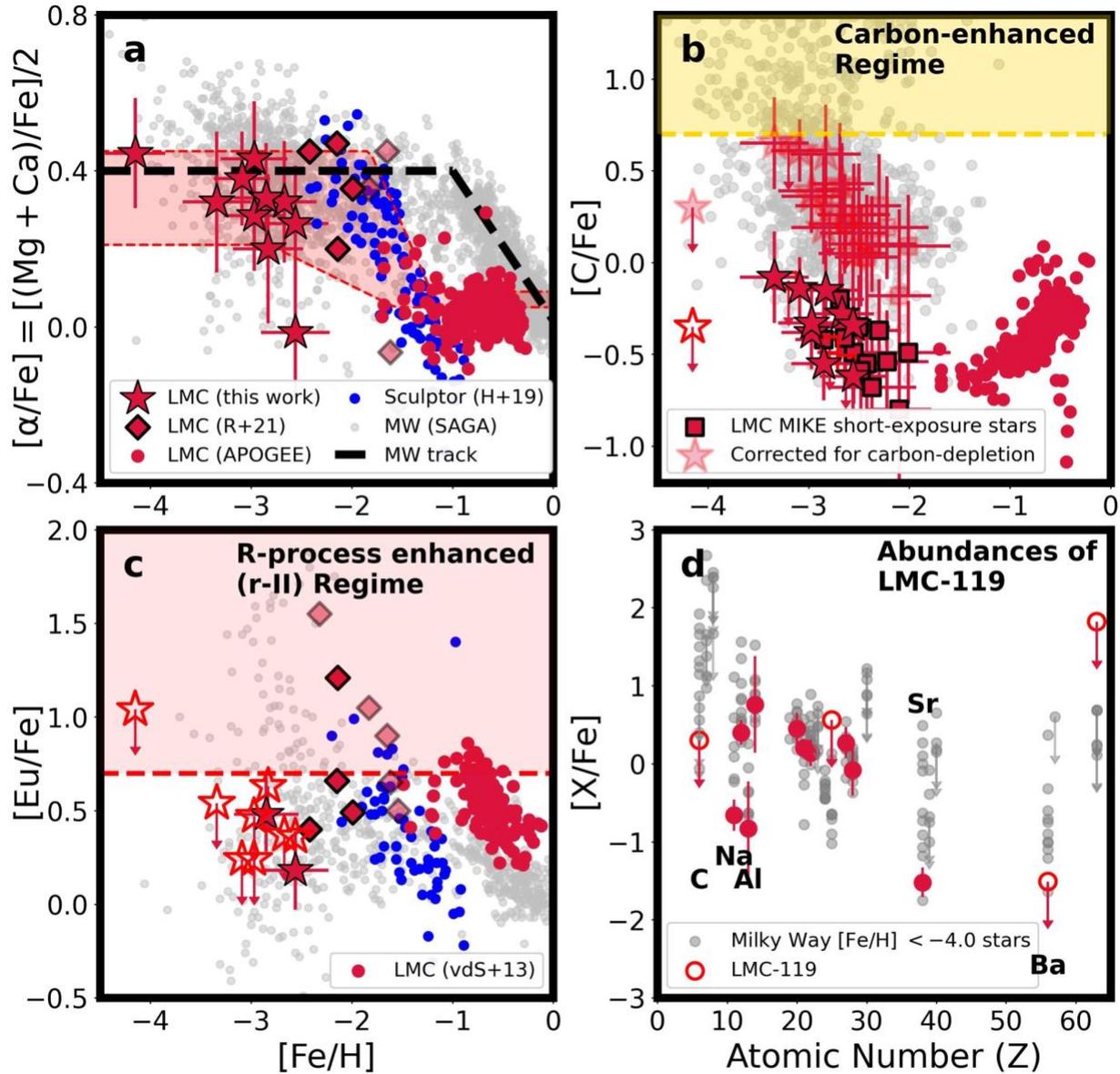

**Figure 2:** Elemental abundance trends of stars in the LMC, compared to the Milky Way and Sculptor, a smaller dwarf galaxy.

**a.** The combined α-element abundance trend for the LMC is shown in red, for the Sculptor dwarf galaxy in blue[35], and for the Milky Way compilation of metal-poor stars in the SAGA database in gray[42]. The α-element track of the Milky Way[33] is shown as a dashed line. The trend seen in the LMC is clearly discrepant from the Milky Way and roughly matches what is seen in Sculptor, a Milky Way dwarf galaxy that is ~3 orders of magnitude less massive in stellar mass than the LMC[43]. Red stars correspond to elemental abundances presented in this study, red diamonds correspond to those from ref. [6], and red circles correspond to those from the APOGEE survey[13,44]. More transparent diamonds correspond to stars in ref. [6] with very cool effective temperatures (< 4300 K) that may be especially susceptible to systematic effects (e.g., non-LTE). The red shaded region is the 2 sigma range of LMC α-knee fits consistent with the data, with parameters listed in the methods. Error bars, as in panels **b-d**, correspond to 1 sigma random uncertainties as derived in the methods section and shown in Table 2, with N corresponding to the number of absorption features for that particular element and star listed in Supplementary Data 2.

**b.** Carbon abundances as a function of [Fe/H] for LMC and Milky Way stars are shown, with the same labeling scheme as in panel **a.** Lighter stars correspond to our MIKE data after applying a correction for the evolutionary state of carbon following ref. [12]. The yellow region corresponds to the regime at which stars are considered carbon-enhanced metal-poor (CEMP) stars. Even after the significant corrections, none of the LMC stars can be considered CEMP stars, whereas 25% of Milky Way halo stars with [Fe/H] < -2.5 are considered CEMP after the correction[12], and ~90% of those with [Fe/H] < -4.0 are CEMP stars[24]. The Milky Way compilation in this panel is from the SAGA and JINAbase compilation in ref. [24].

**c.** Europium (Eu) abundances as a function of metallicity for LMC and Milky Way stars are shown following the labeling scheme in panel **a.**, except the red circles are LMC measurements taken from ref. [45]. The extreme Eu enhancement in ref. [6] is not seen in the lower metallicity stars in our sample, suggesting a sharp onset of the production of this element at [Fe/H] ~ -2.5.

**d.** Chemical abundances of LMC-119 (the lowest metallicity LMC star) as a function of atomic number are shown in red, with abundances from Milky Way halo red giant stars from the SAGA database with [Fe/H] < -4.0 shown in gray. LMC-119 has abundances generally in line with those seen in other ultra metal-poor Milky Way stars, except for carbon and the elements noted in the plot. The carbon abundance is corrected for the evolutionary state of the star following ref. [12], and we applied the same correction to the compilation in the SAGA database.

| Name | RA | DEC | g | T$_{eff}$ (K) | Log(g) | v$_{mic}$ (km/s) | [Fe/H] |
|---|---|---|---|---|---|---|---|
| **LMC-003** | 04:53:37.090 | −64:43:57.55 | 17.05 | 4540±158 | 0.80±0.33 | 2.90±0.28 | −2.97±0.20 |
| **LMC-100** | 04:33:47.361 | −67:37:31.51 | 17.36 | 4612±162 | 1.10±0.32 | 2.63±0.29 | −2.68±0.22 |
| **LMC-104** | 05:18:01.352 | −63:22:26.56 | 16.50 | 4531±155 | 1.05±0.31 | 2.89±0.28 | −2.56±0.21 |
| **LMC-109** | 04:12:01.964 | −66:00:58.36 | 17.01 | 4504±155 | 1.10±0.31 | 2.58±0.28 | −2.85±0.20 |
| **LMC-119** | 05:51:51.026 | −63:56:37.60 | 16.85 | 4315±207 | 0.80±0.33 | 2.66±0.28 | −4.13±0.20 |
| **LMC-124** | 06:08:14.982 | −62:07:23.33 | 16.76 | 4531±160 | 1.10±0.32 | 2.78±0.29 | −2.96±0.21 |
| **LMC-204** | 04:33:20.143 | −68:02:41.87 | 17.30 | 4594±173 | 0.95±0.36 | 3.02±0.34 | −2.85±0.27 |
| **LMC-206** | 05:45:46.652 | −65:46:35.40 | 17.08 | 4720±170 | 1.65±0.34 | 2.56±0.31 | −2.52±0.24 |
| **LMC-207** | 05:32:54.917 | −64:53:09.88 | 17.10 | 4486±172 | 0.80±0.30 | 2.57±0.32 | −3.33±0.25 |
| **LMC-215** | 05:56:37.323 | −66:05:17.85 | 17.16 | 4567±157 | 0.75±0.35 | 2.50±0.27 | −3.10±0.17 |

**Table 1:** Stars in the Large Magellanic Cloud (LMC) observed with the high-resolution Magellan/MIKE spectrograph for detailed element abundance analysis in this work. Stellar parameters and metallicities of our low metallicity LMC stars. g is the SkyMapper magnitude as inferred from the Gaia XP spectra. T$_{eff}$ (K), Log(g), v$_{mic}$, and [Fe/H] list the effective temperature, surface gravity, microturbulence parameter, and metallicity of the stars, with their respective random uncertainties.

| Name | RV (km/s) | [Fe/H] | [C/Fe] | [C/Fe]_c | [Mg/Fe] | [Ca/Fe] | [Eu/Fe] |
|---|---|---|---|---|---|---|---|
| **LMC-003** | 281.5±2.0 | −2.97±0.20 | −0.37±0.20 | 0.36±0.20 | 0.39±0.20 | 0.19±0.20 | <0.25 |
| **LMC-100** | 289.0±2.0 | −2.68±0.22 | −0.29±0.22 | 0.43±0.22 | 0.38±0.22 | 0.25±0.22 | <0.37 |
| **LMC-104** | 224.5±2.0 | −2.56±0.21 | −0.61±0.21 | 0.16±0.21 | 0.34±0.21 | 0.19±0.21 | 0.18±0.21 |
| **LMC-109** | 222.5±2.0 | −2.85±0.20 | −0.55±0.20 | 0.19±0.20 | 0.43±0.20 | 0.23±0.20 | 0.48±0.20 |

| Star | RV | [Fe/H] | [C/Fe] | [C/Fe]_c | [Mg/Fe] | [Ca/Fe] | [Eu/Fe] |
|---|---|---|---|---|---|---|---|
| **LMC-119** | 300.7±2.0 | −4.13±0.20 | <−0.35 | <0.30 | 0.42±0.20 | 0.47±0.20 | <1.04 |
| **LMC-124** | 284.2±2.0 | −2.96±0.21 | −0.34±0.21 | 0.39±0.21 | 0.49±0.21 | 0.36±0.21 | <0.46 |
| **LMC-204** | 252.3±2.0 | −2.85±0.27 | −0.15±0.27 | 0.59±0.27 | 0.31±0.27 | 0.08±0.27 | <0.62 |
| **LMC-206** | 277.2±2.0 | −2.52±0.24 | −0.34±0.24 | 0.05±0.24 | −0.05±0.24 | 0.02±0.24 | <0.36 |
| **LMC-207** | 334.0±2.0 | −3.33±0.25 | −0.12±0.25 | 0.65±0.25 | 0.16±0.26 | 0.48±0.25 | <0.59 |
| **LMC-215** | 299.6±2.0 | −3.10±0.17 | −0.14±0.17 | 0.61±0.17 | 0.48±0.17 | 0.28±0.17 | <0.24 |

**Table 2:** Radial velocities and selected elemental abundances of our low metallicity LMC stars with high-resolution Magellan/MIKE spectra.

RV is the heliocentric radial velocity of each star, [Fe/H] is the metallicity, [C/Fe] is the carbon abundance derived from the spectra, and [C/Fe]_c is the carbon abundance after correcting for the evolutionary state of the star following ref. [12]. [Mg/Fe], [Ca/Fe], and [Eu/Fe] are the magnesium, calcium, and europium abundances, respectively. The uncertainties listed here are random uncertainties; the full chemical abundances of these stars, abundances derived from individual absorption features, and uncertainties are provided in Extended data tables 3 and 4.

**Methods:**

**Target Selection & Observations**

We identified low metallicity members of the Large Magellanic Cloud (LMC) using the third data release (DR3) of the Gaia astrometric mission. Specifically, Gaia DR3 provided low-resolution flux-calibrated spectra for ~200 million stars (hereafter known as XP spectra) [14]. The Gaia mission also provided a python toolkit (GaiaXPy) that allows computing photometry in a set of provided imaging filter passbands through these XP spectra. We used this toolkit to derive photometry of these stars through the SkyMapper u,v,g,i filters, and the narrow-band CaHK filter centered on 393.3nm. Previous work has shown that photometry with these filters can reliably identify stars in the lowest metallicity regimes [8,16,46].

We obtained metallicities for these 200 million XP spectra by matching the inferred SkyMapper u,v,g,i photometry and CaHK photometry from GaiaXPy to a grid of forward-modeled synthetic photometry from synthetic stellar spectra spanning a range of metallicities (-4.0 < [Fe/H] < -0.5) on the red giant branch. This XP metallicity catalog and the details of this implementation will be presented in Mardini, Chiti, and Frebel, in prep., but we describe the relevant details here. This grid of synthetic spectra and photometry was adopted from refs. [9,47], which was generated using the Turbospectrum radiative transfer code [48,49], the VALD linelist supplemented by updated molecular band information [50,51,52,53,54,55,56], and the MARCS model atmospheres [57]. The synthetic photometry was derived from these spectra following refs. [9,47], to generate contours in color-color space of constant metallicity. The location of the XP-based photometry in the SkyMapper (v-g)-0.9*(g-i) vs. g-i space relative to these contours was used to derive a metallicity (see Figure 3 in ref. [9]). The same procedure was applied to derive a metallicity from the CaHK filter, using the color terms (CaHK - g) - 0.9*(g-i) vs (g-i), providing a total of two estimates of the metallicity for each XP spectrum. For completeness, we note that the surface gravity of each star was fixed to be the value derived from comparing its location in the absolute g magnitude vs. g-i space to Dartmouth isochrones [41]. If this process for deriving surface gravities did not converge, stars were assigned a surface gravity of log g=2.0. We used the two metallicity values ([Fe/H]_SkyMapper_XP and [Fe/H]_CaHK_XP) to select candidates.

From this global catalog of XP spectrum-based photometry and metallicities, we selected all stars within 10 degrees of the LMC and applied a number of quality and selection cuts to ensure a pure sample of low metallicity LMC member stars. First, we removed stars in regions of high reddening (E(B-V) > 0.2 in the dustmaps from ref. [40]). Then, we excluded stars which failed to meet quality criteria listed in the Gaia XP spectra paper (RUWE > 1.4, c_star < 3)[14,17]. We retained stars with reasonable precision in their synthetic SkyMapper v and CaHK magnitudes (< 0.2 mags uncertainty). This effectively limited our sample to SkyMapper g < 17.4. After this, stars were selected to have proper motions consistent within 2 sigma of the proper motion distribution of LMC stars[15] and to have locations consistent within 0.1 mags of a [Fe/H] = -2.5, 12 Gyr Dartmouth isochrone [41] placed at a distance modulus of 18.5 [58]. The procedure outlining this selection is shown in Figure 1.

We observed our targets with Magellan/MIKE over three blocks of nights: November 30 & December 1, 2022; December 19 & December 20, 2022; and January 27, 2023. The details of these observations are shown in Extended Data Table 1. The 1".0 slit and 2x2 binning were used for all observations, which yielded wavelength coverage from ~330nm to ~900nm and a resolution of R~28,000 over blue wavelengths and R~22,000 over the red wavelengths[18]. On November 30th, the humidity forced the telescope to close after ~3 hours. Snapshot spectra were obtained for a few LMC stars, selected solely based on [Fe/H]_SkyMapper_XP < -2.5. This initially resulted in a ~20% (1 out of 5 stars; see Extended data Table 1) success rate of finding stars with [Fe/H] < -2.5. On December 1st, we updated this selection to also require [Fe/H]_CaHK_XP < -2.5, with a preference given to targets with lower [Fe/H]_CaHK_XP. This resulted in a 100% success rate of finding stars with [Fe/H] < -2.5 – all six stars that were observed were verified to have such low metallicities from their MIKE spectra (LMC100, LMC104, LMC119, LMC204, LMC206, LMC207). On December 19-20, three additional low metallicity stars were observed following the same selection criteria (LMC109, LMC124, LMC215), and two short-exposure spectra of more metal-rich stars were obtained that passed our selection. The January 27, 2023 observations are discussed in the next paragraph.

Upon initial analysis of these stars, we noticed that none had a strong CH absorption feature at ~431nm, indicating a lack of carbon. The SkyMapper passband has a known selection effect against carbon-enhanced metal-poor stars due to the presence of the CN molecular absorption feature in the passband of the SkyMapper v filter (see Figure 2 in ref. [8]). This feature makes carbon-enhanced metal-poor stars appear artificially metal-rich in this band, increasing the likelihood that they are de-selected for metal-poor star follow-up. A full discussion of this effect is presented in the section "Computation of the carbon-enhanced metal-poor fraction & discussion of selection effects" below, but we briefly describe how we actively compensated for this in our January observations here. For the sample observed on January 27, 2023, we opted to obtain MIKE spectra with short exposures to prioritize deriving solely metallicities and carbon abundances of a larger sample of stars with the same instrumental setup as before. These stars were selected to have [Fe/H]_SkyMapper_XP > -3.0 and [Fe/H]_CaHK_XP < -3.0, to proactively select for any carbon-enhanced metal-poor stars, since the narrower CaHK passband is less sensitive to moderately enhanced CN features ([C/Fe] ~ 0.5; see effect of extreme C-enhancement in Figure 2 in ref. [8]). Twelve stars were observed and they all had [Fe/H] < -2.0 (see Extended Data Table 2). None were carbon-enhanced. We then observed an additional three stars, selected in the same way, with the medium-resolution (R ~ 6000) Magellan/MagE spectrograph[59] with the 0."7 slit on March 16 and March 17, 2023 to derive metallicities and carbon abundances. These stars are also listed in Extended Data Table 2.

We note that given the faintness of our targets, the XP metallicity uncertainties were typically > 0.5 dex making our selections closer to a qualitative cut for low metallicity stars. Due to these large uncertainties arising from source faintness, the metallicities derived from the MIKE spectra show a large scatter around the XP metallicities (systematic offset of 0.27 dex offset, standard deviation of 0.53 dex).

**Chemical abundance analysis of spectra**

Three sets of chemical abundance analysis procedures were used to analyze our sample, corresponding to each of the three subsamples: the long-exposure MIKE spectra for detailed chemical abundances (in Tables 1, 2 and Supplementary Data 1, 22); the short-exposure MIKE data for just metallicities and carbon abundances (in Extended Data Table 2); and the MagE data for metallicities and carbon abundances (also in Extended Data Table 2).

For the sample of long-exposure MIKE spectra, we follow standard stellar chemical abundance analysis methods using ATLAS9 1D model atmospheres[60], the 2017 version of the MOOG radiative transfer code[61,62], and the linelist from ref. [63]. This analysis was performed in the Spectroscopy Made Harder (SMHR) python wrapper [64], generally following the procedures in refs. [19,65] which we briefly describe here. We fit equivalent widths to derive chemical abundances for each line by fitting a Gaussian profile after local continuum normalization. Initial guesses for the stellar parameters (effective temperature, microturbulence, surface gravity) were based on the location of these stars on the LMC isochrone. Then, these stellar parameters were iteratively adjusted until the Fe I abundances showed no trend with excitation potential and reduced equivalent width. Additionally, the surface gravity was adjusted until the Fe I and Fe II average abundance was in agreement. Then, the temperature correction from ref. [19] was applied to bring the temperature scale in line with photometric values, and the microturbulence and surface gravities were again adjusted until the Fe I abundances showed no trend with reduced equivalent width, and the Fe I and Fe II average abundances agreed. The chemical abundances from molecular bands or dense regions of absorption features were inferred by fitting synthetic spectra to those regions of the spectrum and varying the abundance of the element of interest.

The final chemical abundances and uncertainties for the long-exposure sample are computed exactly following ref. [65], which we outline here. The final chemical abundance is taken as the mean abundance from individual absorption features of that element. For elements with > 9 absorption features, the standard deviation of the abundances is taken as the random uncertainty. For elements with 2 to 9 absorption features, the random uncertainties are computed by multiplying the range of abundances by the k-statistic [66]. For elements with 1 absorption feature, the uncertainty is taken as the abundance uncertainty inferred from the fit parameters to the feature. For Si and Al, the absorption features we use are relatively insensitive to the abundances for these stellar parameters, so an uncertainty floor of 0.55 dex is added to those particular abundances. The random uncertainties on the stellar parameters are taken to correspond to the allowable range from the scatter in the Fe I line trends with excitation potential and reduced equivalent width, and the Fe I and Fe II abundances. These random uncertainties are added in quadrature to the systematic uncertainty in the stellar parameters, which are taken to be fiducial values of 150 K in effective temperature, 0.3 dex in surface gravity, and 0.2 km/s in microturbulence following ref. [65]. The shifts from varying the stellar parameters by their uncertainties and re-deriving chemical abundances are assumed to be the systematic uncertainties on the chemical abundances. These are added to the random uncertainties in the chemical abundances in quadrature to derive final chemical abundance uncertainties. For features with no detected lines, three sigma upper limits from spectrum synthesis are reported. The full chemical abundances and associated uncertainties of this sample is provided in Supplementary Data 1 and 2, and the suite of chemical abundances are plotted in Extended data Figure 3.

For the short-exposure MIKE data, the same procedure as above was applied with a few notable exceptions. First, the stellar parameters were fixed to the value as inferred by the location of these stars on a 12 Gyr, [Fe/H] = -2.0 Dartmouth isochrone [41]. This was done since too few Fe I lines were detected in each spectra to robustly constrain the stellar parameters. The uncertainties in these stellar parameters are taken to be the fiducial uncertainties above. Second, the linelist for the Fe I lines was chosen to be from the RPA2k linelist provided with the Payne4MIKE code, which includes a clean selection of lines from stars based on stars analyzed by the R-process Alliance[67,68,69,70]. Third, the carbon abundance was only derived from the CH absorption band at 431 nm. The iron abundance and carbon abundances for these stars are reported in Extended Data Table 2.

For the MagE spectra, the stellar parameters were derived following the same procedure as for the short-exposure MIKE data. The metallicities were derived from the Ca H and K lines using the well-established KP calibration [71,72], implemented following refs. [73,74]. The carbon abundances were derived using spectrum synthesis, again following ref. [74], which replicated this analysis on similar low metallicity stars in the Sagittarius dwarf spheroidal galaxy.

**Orbit integration to confirm membership**

We note that while all of our stars fall within the proper motion and radial velocity criteria for membership of the LMC following ref. [13], a few stars have proper motion vectors antiparallel to the bulk motion of nearby LMC members in the center-of-mass reference frame of the LMC. To ensure that our low metallicity stars are in fact members of the LMC, we integrate their orbits backward in time to check whether they are gravitationally bound to the LMC. We use the public code in ref. [20] which includes the Milky Way, LMC, and Sagittarius dwarf galaxy, and assuming the LMC stars are at a distance of 50 kpc. As shown in panel (a) of Extended data Figure 2, these stars have been bound to the LMC, confirming that they are likely members.

**Analysis of the distance between the LMC and MW at high redshifts**

We estimate the distance between the LMC and the MW by analyzing a suite of cosmological simulations from the *Caterpillar* Project [75]. The *Caterpillar* Project includes 32 dark-matter-only zoom-in Milky Way-mass halo simulations. Among the satellite dwarfs in these simulations, we selected the most "LMC-like" satellites by identifying, for each simulation, the most recently accreted or infalling halo with peak mass ratio between 1:10 to 1:4 compared to the MW-mass halo. We also require that they reach peak mass and begin infalling into the "MW" at z < 1. This produces 18 LMC-like halos. After identifying the most LMC-like halos, we trace both their history and the history of the MW-mass halo through the simulation back to redshifts z ~ 6 to z ~ 9, the ages we expect the halo to be forming stars with [Fe/H] = -4.0 to [Fe/H] = -2.5, following refs. [76,77]. We calculate the physical distance between the "LMC" and the "MW" between these points in time and find that the distances range from 1 and 3 comoving Mpc. This isolates the early LMC from ejecta from the first stars that formed in the early

Milky Way[78], and places the LMC at the scale of large-scale structure variations in the early universe[22].

For completeness, we note that there are scenarios in the literature in which the LMC is on second-passage (e.g., [79] and as discussed for an unlikely low-mass LMC in ref.[80]). In the scenario in ref[79], the previous pericentric passage of the LMC could have occurred 5-10 Gyr ago, but at a large distance of > 100 kpc. In Figure 4 of ref. [79], the distance between the LMC and the Milky Way in this second-infall scenario ~11 Gyr ago appears to still be quite far (>~400 kpc). This is still sufficiently distant for the LMC ecosystem to be shielded from ejecta of the first stars formed in the Milky Way ecosystem[81,82].

**Supernova yield fitting**

The most metal-poor star in our sample, LMC-119, is plausibly a second-generation star due to its low metallicity[10,11]. Consequently, we fit its chemical abundance pattern using supernova yield models of the first stars from ref. [23] following the procedure in ref. [83]. Briefly, we use fits from these models to estimate the distribution of preferred progenitor star masses, supernova explosion energy, and dilution mass following a chi-squared fitting procedure to the chemical abundance pattern of LMC-119. We find that a broad range of masses (10Msun to 50Msun), explosion energies (up to $4\times10^{51}$ergs), and dilution masses ($10^5$Msun to $10^{6.5}$Msun) are consistent with the abundance pattern of LMC-119. Note that only C, Na, Mg, Ca, Ti, Fe, Co, Ni are included in this analysis due to a combination of their relatively lower abundance uncertainties/upper limits, and somewhat less sensitivity to NLTE effects than other elemental abundances. The results of this fitting procedure are shown in Extended data Figure 4.

**α-knee fitting**

Abundances of **α**-elements (e.g., Mg, Ca) as a function of [Fe/H] trace the efficiency of star formation. This is because core-collapse supernovae dominate early metal production and over-produce these elements ([α/Fe] ~ 0.4[84]). Type Ia supernovae occur after some delay time[85] and produce only traces of α-elements[86] but significant amounts of iron-peak elements (e.g., Fe, Ni, Mn). Consequently, the [Fe/H] at which [α/Fe] starts to decline maps onto the early star formation efficiency. We model the location of the downturn in the α-element abundances (the α-"knee") as a function of [Fe/H] largely following ref. [87], which we outline here. First, we take [α/Fe] = ([Mg/Fe] + [Ca/Fe])/2, as these two α-elements are typically those with the smallest uncertainties. We model the trend of [α/Fe] vs. [Fe/H] as a piecewise linear model with four parameters: [α/Fe](high), which is the average [α/Fe] value at the low metallicity end; [Fe/H](low), which indicates the metallicity at which [α/Fe] starts declining; [Fe/H](high), which indicates the metallicity at which [α/Fe] stops declining; and [α/Fe](low), which is the constant [α/Fe] value at the high metallicity end.

We fit the above model to a compilation of LMC data from refs. [6,13,44] and our sample of low metallicity LMC stars. The fitting was performed using the scipy curve_fit function, allowing [α/Fe](high) and [α/Fe](low) to vary by 0.2 dex around the mean [α/Fe] of stars with [Fe/H] < -3.0 and [Fe/H] > -1.0, respectively. Additionally, [Fe/H](high) was allowed to vary between -2.0 and -0.5, and [Fe/H](low) was allowed to vary between -3.0 and -1.0. Then, the LMC data compilation was re-sampled 10000 times based on the chemical abundance uncertainties and the same parameters were re-derived each time. The standard deviation of the resulting parameter distribution is taken as the random uncertainty in the parameter values. This random uncertainty in the model fit is added to the errors inferred from the covariance matrix returned by curve_fit in quadrature to derive the final uncertainty value on each of these parameters. Notably, with [Fe/H](low), we constrain the location of the α-"knee" to be at [Fe/H] < -1.82, significantly lower than what is expected ([Fe/H] ~ -1.2[13]) for a galaxy the mass of the LMC [34]. Visually, this low downturn is also evident in panel (a) of Figure 2 and in the individual [Ca/Fe], [Mg/Fe] trends when combined with literature samples. The fits to the parameters are: [α/Fe](high) = 0.33 +/- 0.06, [Fe/H](high) = -1.39 +/- 0.09, [α/Fe](low) = 0.07 +/- 0.01, [Fe/H](low) = -2.32 +/- 0.25. We emphasize that we report the latter fit in terms of its 2sigma upper limit when discussing the α-"knee", due to the visually highly uncertain location of the specific downturn in [α/Fe] (see panel a of Figure 2). Excluding the cool stars (< 4300 K) in ref. [6] shifts the location of the above fits less than 0.1 dex. Fully excluding the stars in ref. [6] leads to an artificially low α-"knee" of [Fe/H] = -2.73, since the stars in ref. [6] largely populate metallicity range between [Fe/H] > -2.5 and [Fe/H] < -2.0. Excluding our new low metallicity sample leads to a fit that does not converge.

We note that we use this simple piecewise linear model only for insights on the approximate initial downturn of the α-knee. This model is likely insufficient to explain the full morphology of the trend. Accordingly, the exact parameters of this model should not be assumed to fully represent the LMC α-track. A more complete description of the LMC α-element evolution can be obtained for [Fe/H] > -2.2 in the b-spline fits in Figure 14 of ref. [13].

**Computation of the carbon-enhanced metal-poor fraction & discussion of selection effects**

The carbon-enhanced metal-poor fraction for our sample was taken as the fraction of stars with [C/Fe]_c > 0.7 divided by the total number of stars, which allows a consistent comparison to Milky Way carbon-enhanced fractions reported in the literature[12]. Note that the carbon abundances that were derived from the spectra are corrected for the evolutionary state of the star following ref. [12]. This is because stars deplete carbon in their photospheres as they ascend the red giant branch, so this correction is applied to recover their natal abundance of carbon. Since all of our LMC stars are well-ascended on the red giant branch, their carbon corrections are often significant (~0.5 to ~0.7 dex) as seen in Table 2 and Extended Data Table 2.

Given the low carbon abundances of the low metallicity stars that we detected in the LMC, we performed several checks to ensure that biased target selection could not reasonably explain carbon deficiency. In particular, these stars were flagged as low metallicity candidates due to their brightness in the metallicity-sensitive filter covering the CaII H and K absorption feature at 393nm. There is a prominent CN absorption band blueward of this feature (at ~385nm) that can make stars appear faint (and thus, more metal-rich) if the carbon abundance is sufficiently high. However, the strength of this spectral feature depends on multiple quantities – the effective temperature of the star, carbon abundance, metallicity, and evolutionary state which affects the surface carbon and nitrogen abundance. The strength of the band increases with decreasing temperature, increasing metallicity, and increasing carbon abundance. Additionally, the carbon-enhanced threshold (corrected [C/Fe] > 0.7) for our stars would generally be met already when the inferred morphology of the CN band is fairly weak (e.g., corresponding to uncorrected [C/Fe] ~ 0) due to the significant corrections for carbon depletion for stars on the red giant branch.

Accordingly, the effect of our selection function must be assessed using a sample of stars spanning similar metallicity, effective temperature, and surface gravity. We queried the SAGA database of metal-poor stars for stars[42] with effective temperatures <4750K and log g < 2.0, matching our sample of LMC stars. We then cross-matched this compilation with our XP catalog of metallicities. We replicated our selection function for the LMC stars on this cross-matched catalog. Specifically, all 10 of our long-exposure LMC stars had [Fe/H]_CaHK_XP < -2.8 and [Fe/H]_SkyMapper_XP < -3.0. We computed the fraction of stars below [Fe/H] < -2.5 in SAGA that are above various uncorrected [C/Fe] thresholds before and after applying these selection criteria. Notably, we find no meaningful difference in the recovered fraction of stars with [C/Fe] > 0.0 before (N=23, fraction=0.36) and after (N=11, fraction=0.34) applying our selection, implying that we ought to have found stars with uncorrected [C/Fe] > 0 in an unbiased manner in the metallicity regime considered for this study. None of the LMC stars have uncorrected [C/Fe] > 0.0, as provided in Table 2 or Extended data Table 2. We note that none of the three most carbon-enhanced stars in SAGA (all with uncorrected [C/Fe] > 0.65) in this metallicity regime are recovered by our selection. Consequently, we are biased against the most drastically carbon-enhanced stars in the LMC which is consistent with previous work showing a selection against significant carbon-enhancement by these photometric filters [8,9,46,88]. However, as shown by the SAGA analysis, we see no evidence for a bias against moderately carbon-enhanced stars for the stellar parameter and metallicity regime in this study, exactly in the carbon regime where we do not readily detect these stars in the LMC. An uncorrected [C/Fe] > 0 roughly maps to [C/Fe] > 0.7 after applying carbon corrections for typical stellar parameters of our sample. We note that the short-exposure MIKE and MagE sample would be less susceptible to any biases discussed here since they were chosen to have [Fe/H]_SkyMapper_XP > -3.0 (see Target Selection & Observations).

We note that, generally, the most carbon-enhanced stars (e.g., uncorrected [C/Fe] > 1.0) can also appear meaningfully redder than the adopted isochrone for target selection, leading to a potential exclusion from our selection in panel (b) of Figure 1. For example, ref. [89] recently identified a very metal-poor, extremely carbon-enhanced ([Fe/H] = -2.5, [C/Fe] ~ 2.5) star in Reticulum II that was 0.25 mag redder than the isochrone overplotted in their Figure 1. We likely exclude such stars from selection in the LMC due to the 0.1 mag range of our isochrone selection. However, as also seen in Figure 1 of ref. [89], the other stars in Reticulum II track the isochrone closely and have corrected carbon abundances spanning [C/Fe] = 0.16 to [C/Fe] = 1.29 [90]. Consequently, our isochrone selection principally excludes extremely carbon-enhanced stars in the LMC but ought to have still selected stars well beyond the CEMP threshold ([C/Fe] > 0.7), as shown in and spanning the y-axis of panel (b) in Figure 2. Such extremely carbon-enhanced stars, when they show no enhancement in elements produced by the slow neutron capture process (s-process), are known as Group III stars [92]. We would not identify such stars in our selection, while still being sensitive to so-called Group II carbon-enhanced stars which show a less drastic carbon-enhancement[92] still above the [C/Fe] > 0.7 threshold.

We also test to see whether our sample indicates differences in the mean carbon abundance vs. [Fe/H] relative to the Milky, as opposed to just a differing fraction of stars that are carbon-enhanced ([C/Fe] > 0.7). We divide our LMC sample into bins of 0.5 dex in [Fe/H] from -2.0 to -3.5 and compute the mean [C/Fe] in each of these to compare to the Milky Way halo. We find that the mean carbon abundance of the LMC sample is consistent with the Milky Way halo trend (ranging from 0.03 dex to 0.18 dex below the MW median values), despite the lack of CEMP stars in our LMC sample. This analysis, however, is heavily caveated by the small number of LMC stars (2 to 11 stars) in each metallicity bin.

**Data availability** The velocities and chemical abundances derived from the long-exposure MIKE spectra in this study are presented in Table 2, and Supplementary Data 1. Abundance measurements for individual absorption features in these stars are provided in Supplementary Data 2. We report short-exposure MIKE and MagE metallicities and carbon abundances in Extended Data Table 2. The stellar spectra of these stars are available from the corresponding author upon request. The proper motions of these stars are available from the Gaia data archive (https://gea.esac.esa.int/archive/). The data tables will be posted in machine-readable format at Zenodo DOI 10.5281/zenodo.10032360 upon publication[93].

**Code availability** The stellar synthesis code MOOG that was used to analyze this data can be retrieved from https://github.com/alexji/moog17scat. The analysis package SMHR that wraps around MOOG can be retrieved from https://github.com/andycasey/smhr. The orbit integration code that includes the Milky Way, LMC and Sagittarius can be retrieved from ref. [26]. The Payne4MIKE code used to analyze the short-exposure MIKE spectra can be retrieved from https://github.com/tingyuansen/Payne4MIKE. The chemical abundance analysis of the two MagE spectra was performed using the authors' implementations of published techniques and are straightforward to reproduce from the publications, but are available from the corresponding author upon request.

**Methods references:**

**Acknowledgements** Our data were gathered using the 6.5-m Magellan Baade telescope located at Las Campanas Observatory, Chile. A.C. thanks Alex Drlica-Wagner, Knut Olsen, David Nidever, Guy Stringfellow, Will Cerny, Kim Venn, Anke Arentsen, Vini Placco, Ian Roederer, Piyush Sharda, and Andrey Kravtsov for helpful discussions, and R.P. for their support. A.C. also thanks Vini Placco for providing a compilation of corrected carbon abundances of metal-poor Milky Way stars. This work benefited from the KICP/UChicago Gaia DR3 sprint, and made use of NASAs Astrophysics Data System Bibliographic Services, the SIMBAD database, operated at CDS, Strasbourg, France and the open-source python libraries numpy, scipy, matplotlib, and astropy. A.C. is supported by a Brinson Prize Fellowship at the University of Chicago/KICP. G.L. acknowledges FAPESP (procs. 2021/10429-0 and 2022/07301-5). A.P.J. acknowledges support by the U.S. National Science Foundation under grants AST-2206264 and AST-2307599.

This work has made use of data from the European Space Agency (ESA) mission Gaia (https://www. cosmos.esa.int/gaia), processed by the Gaia Data Processing and Analysis Consortium (DPAC, https://www. cosmos.esa.int/web/gaia/dpac/consortium). Funding for the DPAC has been provided by national institutions, in particular the institutions participating in the Gaia Multilateral Agreement.

The national facility capability for SkyMapper has been funded through ARC LIEF grant LE130100104 from the Australian Research Council, awarded to the University of Sydney, the Australian National University, Swinburne University of Technology, the University of Queensland, the University of Western Australia, the University of Melbourne, Curtin University of Technology, Monash University and the Australian Astronomical Observatory. SkyMapper is owned and operated by The Australian National University's Research School of Astronomy and Astrophysics.

**Author Contributions** A.C. designed the technique for generating the Gaia XP metallicity catalog used in this work, selected candidates for the observations, and led the MIKE observations, analysis, interpretation, and paper writing; M.M. generated the Gaia XP metallicity catalog, and assisted with the MIKE observations, analysis, and interpretation; A.P.J assisted with the analysis, interpretation, and paper writing; A.F. assisted with the MIKE observations, interpretation, and paper writing; G.L. assisted with the analysis, interpretation, and paper writing; H.R. provided existing MIKE observations of LMC stars and assisted with the interpretation; P.F. generated the catalog of synthetic Gaia XP photometry; K.B. and H.D.A. led the analysis of the distance of the LMC from the MW when it produced low metallicity stars; T.S.L and J.D.S led and assisted with the MagE observations; all authors provided feedback on the manuscript before submission.

**Competing Interests** The authors declare no competing financial or non-financial interests.

**Additional Information** Correspondence & request for materials should be sent to Anirudh Chiti (achiti@uchicago.edu).

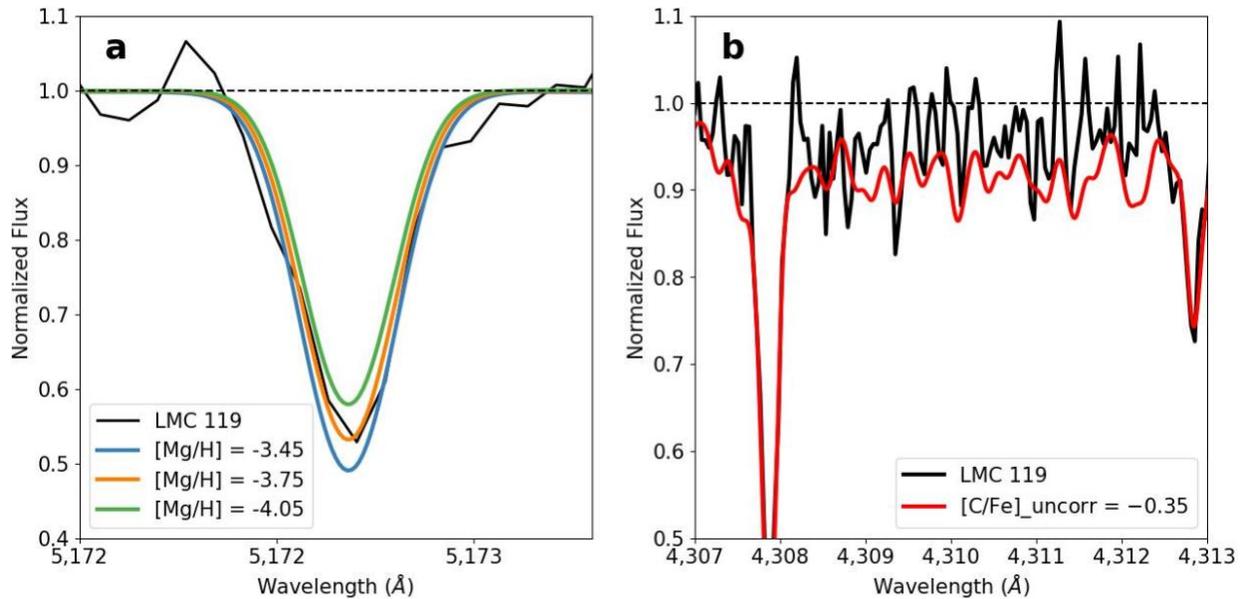

**Extended data Figure 1:** Selected element absorption features in the spectrum of star LMC-119, with model-generated spectra at different abundances shown for comparison.
**a.** A magnesium absorption line at 517.2 nm in the spectrum of LMC-119 (black) with synthetic spectra of various magnesium abundances overplotted. The range of best fitting magnesium abundances matches the magnesium abundance of the star in Table 2. Note that the synthetic spectrum has the same stellar parameters as LMC-119.
**b.** The CH absorption band at ~431 nm in the spectrum of LMC-119, with a synthetic spectrum of [C/Fe] = -0.35 overplotted. Note the lack of a strong absorption feature in LMC-119.

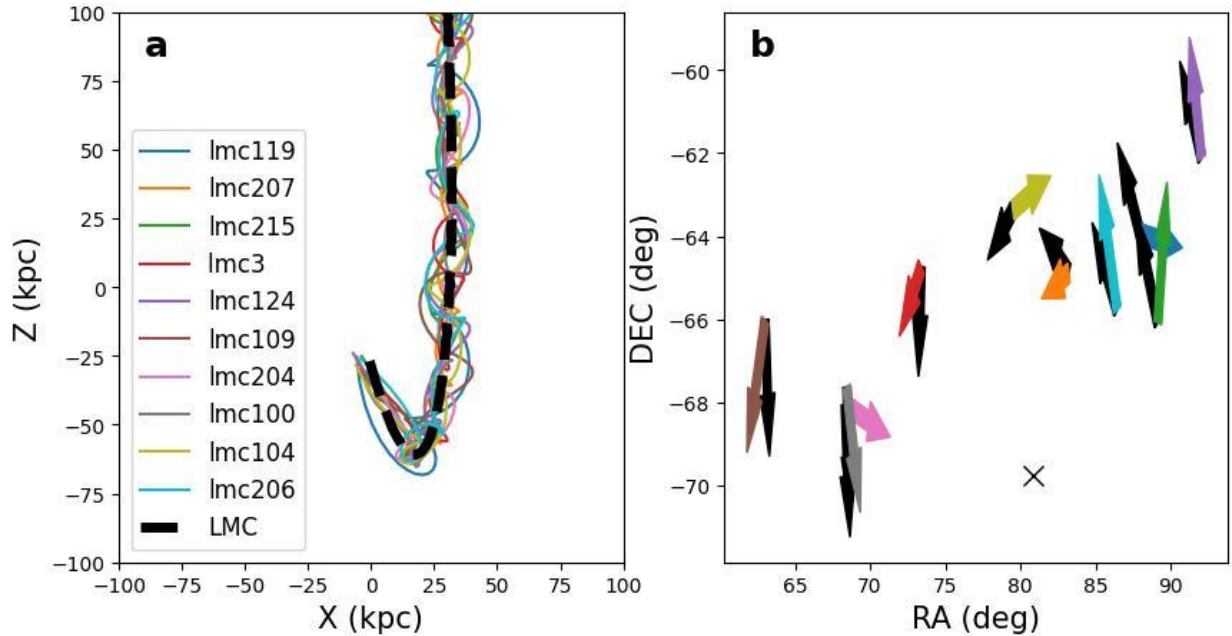

**Extended data Figure 2:** Orbit and proper motion analyses of metal-poor LMC stars in this study.

**a.** The orbit of the LMC and stars observed in this study in Galactocentric X and Z coordinates, when integrated backwards in a potential that includes the Milky Way and the LMC [20]. The LMC is shown as a thick dashed line, and stars in Table 1 are overplotted. Note that the stars remain bound to the LMC, indicating that they are gravitationally bound members.

**b.** Proper motion vectors with respect to the center of mass motion of the LMC. The proper motions of the stars observed in our study are shown as colored arrows, with the color scheme corresponding to panel (a). The black arrows indicate the average proper motion vector of LMC stars within 0.5° of their spatial location. Note that two of the LMC low metallicity stars are counter-rotating on the plane of the sky with respect to the bulk motion of LMC members, in the center-of-mass frame of the LMC. The center of the LMC is marked by a black cross.

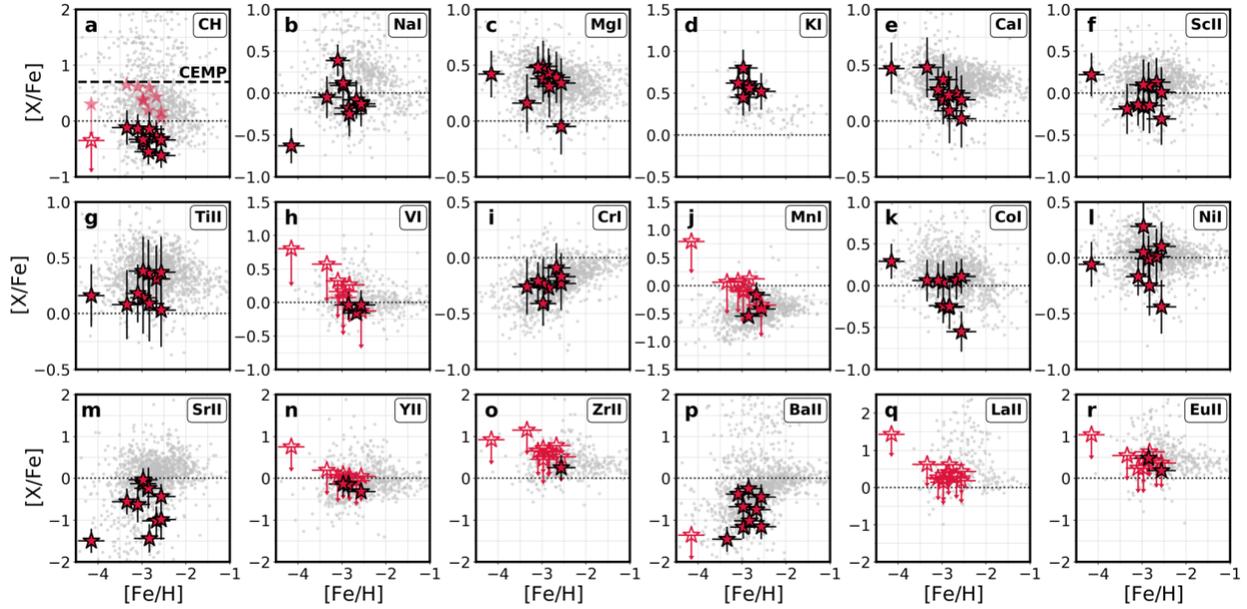

**Extended data Figure 3:** Complete elemental-abundance inventory for the low-metallicity LMC stars with long-exposure, high-resolution Magellan/MIKE spectra. **a.-r.** All panels show [X/Fe] versus [Fe/H], where 'X' is an identified species according to the text boxes in the top right corner of each panel. Symbols with black edges represent detections, while open ones are upper limits. The background gray points are Milky Way stars compiled from the SAGA database. Dotted lines at [X/Fe] = 0 in every panel marks the solar level. For the carbon panel on the upper left, the dashed black line indicates the [C/Fe] = +0.7 boundary for the CEMP definition. Also, transparent symbols indicate corrected values for [C/Fe] as in ref. [12]. Error bars correspond to 1 sigma random uncertainties, as derived following the methods section and listed in Supplementary Data 1, with N corresponding to the number of absorption features for that particular element and star listed in Supplementary Data 2.

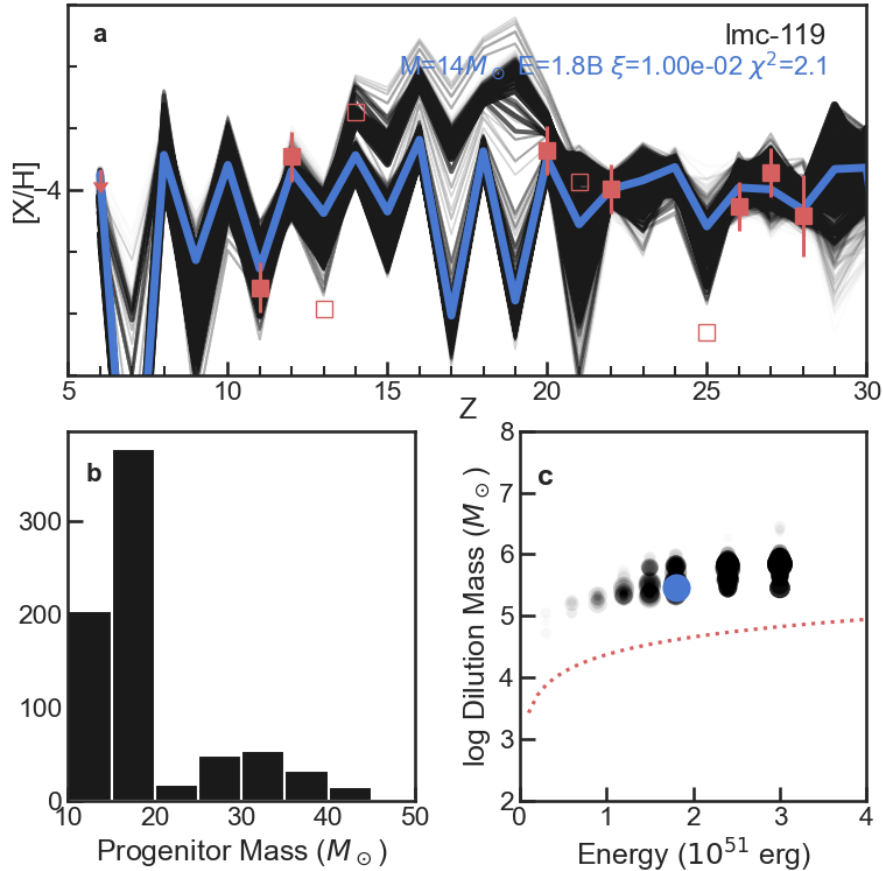

**Extended data Figure 4:** Fitting modeled yields of various supernovae to the chemical abundances of LMC-119.

**a.** Individual fits of supernova yield models from ref. [23] are overplotted as black lines, with chemical abundances shown as red squares. The fitting procedure follows what is described in the methods section. The best fitting yield model and parameters are shown in blue. Note that a range of models are compatible with the chemical abundance pattern of LMC-119. Error bars in the elemental abundances correspond to 1 sigma random uncertainties, as derived following the methods section and listed in Supplementary Data 1, with N corresponding to the number of absorption features for that particular element and star listed in Supplementary Data 2.

**b.** Histogram of masses of progenitors in the supernova yield models consistent with the abundances of LMC-119.

**c.** Distribution of energies and dilution masses of supernova yield models consistent with the abundances of LMC-119. The red curve indicates the minimum dilution mass for a given energy, following ref. [83].

| Name | RA | DEC | g | t_exp (min) | Date | Instrument |
|---|---|---|---|---|---|---|
| LMC-003 | 04:53:37.090 | -64:43:57.55 | 17.05 | 100 | 30 Nov, 2022 | MIKE |
| LMC-100 | 04:33:47.361 | -67:37:31.51 | 17.36 | 25; 80 | 1 Dec, 2022; 19 Dec, 2022 | MIKE |
| LMC-104 | 05:18:01.352 | -63:22:26.56 | 16.50 | 15; 30; 30 | 1 Dec, 2022; 19 Dec, 2022; 20 Dec, 2022 | MIKE |
| LMC-109 | 04:12:01.964 | -66:00:58.36 | 17.01 | 20; 40 | 19 Dec, 2022; 20 Dec, 2022 | MIKE |
| LMC-119 | 05:51:51.026 | -63:56:37.60 | 16.85 | 105 | 1 Dec, 2022 | MIKE |
| LMC-124 | 06:08:14.982 | -62:07:23.33 | 16.76 | 60 | 20 Dec, 2022 | MIKE |
| LMC-204 | 04:33:20.143 | -68:02:41.87 | 17.30 | 47 | 1 Dec, 2022 | MIKE |
| LMC-206 | 05:45:46.652 | -65:46:35.40 | 17.08 | 110 | 1 Dec, 2022 | MIKE |
| LMC-207 | 05:32:54.917 | -64:53:09.88 | 17.10 | 115 | 1 Dec, 2022 | MIKE |
| LMC-215 | 05:56:37.323 | -66:05:17.85 | 17.16 | 105 | 19 Dec, 2022 | MIKE |
| LMC-319 | 05:38:55.896 | -64:21:50.52 | 17.33 | 15 | 17 Mar, 2023 | MagE |
| LMC-323 | 06:17:10.604 | -73:59:35.81 | 17.46 | 10 | 16 Mar, 2023 | MagE |
| LMC-400 | 05:08:33.774 | -72:25:27.55 | 16.82 | 15 | 27 Jan, 2023 | MIKE |
| LMC-402 | 04:28:27.512 | -69:22:08.36 | 17.21 | 20 | 27 Jan, 2023 | MIKE |
| LMC-403 | 05:27:13.519 | -65:11:46.17 | 17.06 | 15 | 27 Jan, 2023 | MIKE |
| LMC-404 | 04:35:40.896 | -66:09:29.18 | 17.45 | 15 | 17 Mar, 2023 | MagE |
| LMC-405 | 04:54:16.665 | -64:50:01.06 | 17.38 | 25 | 27 Jan, 2023 | MIKE |
| LMC-406 | 05:15:58.050 | -64:49:18.10 | 16.87 | 12.5 | 27 Jan, 2023 | MIKE |
| LMC-407 | 04:55:47.448 | -64:12:45.29 | 17.33 | 20 | 27 Jan, 2023 | MIKE |

| LMC-409 | 05:55:03.971 | -63:54:43.19 | 16.95 | 15 | 27 Jan, 2023 | MIKE |
| LMC-410 | 05:42:10.807 | -63:53:20.39 | 17.26 | 20 | 27 Jan, 2023 | MIKE |
| LMC-411 | 05:29:44.001 | -63:54:12.14 | 16.69 | 12.5 | 27 Jan, 2023 | MIKE |
| LMC-413 | 05:37:05.212 | -64:34:42.64 | 16.86 | 12.5 | 27 Jan, 2023 | MIKE |
| LMC-414 | 05:27:13.879 | -61:48:50.35 | 17.16 | 20 | 27 Jan, 2023 | MIKE |
| LMC-416 | 06:45:46.181 | -66:29:56.95 | 16.84 | 12.5 | 27 Jan, 2023 | MIKE |
| LMC-007 | 05:44:58.823 | -72:41:45.25 | 17.04 | 20 | 30 Nov, 2022 | MIKE |
| LMC-008 | 06:27:25.744 | -67:01:18.05 | 16.90 | 20 | 30 Nov, 2022 | MIKE |
| LMC-101 | 04:42:26.547 | -65:45:50.31 | 17.07 | 20 | 30 Nov, 2022 | MIKE |
| LMC-116 | 05:46:35.167 | -64:20:32.05 | 17.18 | 15 | 30 Nov, 2022 | MIKE |
| LMC-202 | 05:04:52.769 | -72:20:42.85 | 17.42 | 25 | 19 Dec, 2022 | MIKE |
| LMC-217 | 05:44:00.168 | -62:10:47.41 | 17.08 | 25 | 19 Dec, 2022 | MIKE |

**Extended Data Table 1:** Summary of our observations of stars in the Large Magellanic Cloud.

The Right Ascension (RA) and Declination (DEC) columns indicate the position of the star in the sky. g is the SkyMapper magnitude as inferred from the Gaia XP spectra. t_exp is the total exposure time in minutes, with exposures on separate dates separated by a semicolon. The dates list the night of observation, with multiple nights of observation separated by a semicolon. The last column lists the instrument used for the observation.

| Name | Teff (K) | Log(g) | v_mic (km/s) | [Fe/H] | [C/Fe] | [C/Fe]_c |
| --- | --- | --- | --- | --- | --- | --- |

| Star | Teff | logg | v_mic | [Fe/H] | [C/Fe] | [C/Fe]_c |
|---|---|---|---|---|---|---|
| LMC-319 | 4631±150 | 1.29±0.30 | 2.05±0.2 | -2.69±0.24 | -0.20±0.36 | 0.41±0.36 |
| LMC-323 | 4609±150 | 1.25±0.30 | 2.08±0.2 | -2.10±0.32 | -0.80±0.47 | -0.18±0.47 |
| LMC-400 | 4422±150 | 0.9±0.30 | 2.26±0.2 | -2.85±0.28 | -0.42±0.27 | 0.31±0.27 |
| LMC-402 | 4545±150 | 1.12±0.30 | 2.14±0.2 | -3.20±0.25 | <-0.10 | <0.64 |
| LMC-403 | 4504±150 | 1.05±0.30 | 2.18±0.2 | -2.55±0.36 | -0.49±0.28 | 0.26±0.28 |
| LMC-404 | 4676±150 | 1.38±0.30 | 2.01±0.2 | -2.42±0.32 | -0.55±0.34 | 0.03±0.34 |
| LMC-405 | 4653±150 | 1.33±0.30 | 2.04±0.2 | -2.50±0.31 | -0.35±0.28 | 0.26±0.28 |
| LMC-406 | 4463±150 | 0.97±0.30 | 2.22±0.2 | -2.22±0.41 | -0.54±0.28 | 0.19±0.28 |
| LMC-407 | 4631±150 | 1.29±0.30 | 2.05±0.2 | -2.01±0.42 | -0.49±0.28 | 0.09±0.28 |
| LMC-409 | 4504±150 | 1.04±0.30 | 2.18±0.2 | -2.63±0.21 | <-0.55 | <0.21 |
| LMC-410 | 4609±150 | 1.25±0.30 | 2.07±0.2 | -2.64±0.36 | -0.41±0.27 | 0.21±0.27 |
| LMC-411 | 4382±150 | 0.83±0.30 | 2.3±0.2 | -2.79±0.17 | <-0.45 | <0.28 |
| LMC-413 | 4463±150 | 0.97±0.30 | 2.22±0.2 | -2.37±0.44 | -0.68±0.26 | 0.10±0.26 |
| LMC-414 | 4587±150 | 1.2±0.30 | 2.1±0.2 | -2.30±0.22 | -0.37±0.28 | 0.31±0.28 |
| LMC-416 | 4443±150 | 0.94±0.30 | 2.24±0.2 | -2.48±0.22 | -0.59±0.27 | 0.03±0.27 |

**Extended Data Table 2:** Properties of low metallicity LMC stars observed for short-durations on MIKE and MagE to solely derive metallicities and carbon abundances. Derived parameters of low metallicity LMC stars observed for short durations in Extended Data Table 1. The name of the star is followed by its derived effective temperature in Kelvin (Teff), surface gravity (logg), and microturbulence in km/s (v_mic). The metallicity ([Fe/H]), carbon abundance derived from the spectrum ([C/Fe]), and carbon abundance corrected for the evolutionary state of the star ([C/Fe]_c) following ref. [12] are listed.

**Supplementary Data 1:** Complete elemental abundances and associated uncertainties from long-exposure, high-resolution Magellan/MIKE spectra (link).

Columns are: the star name; the atomic number and ionization state of the element; the number of features used (N); the solar abundance (Solar); the absolute abundance (Logeps); the chemical abundance scaled by the solar abundance relative to hydrogen ([X/H]); the ratio with respect to the iron abundance ([X/Fe]); the random uncertainty ([X/H]_err) and an upper-limit flag(ul); errors from propagating the uncertainties in each stellar parameter ([X/H]_errteff, [X/H]_errlogg, [X/H]_errvt); the cumulative uncertainty from stellar parameters ([X/H]_errsys); and the total uncertainty ([X/H]_errtot). Following this are the same columns but with respect to iron ([X/Fe]_errteff, [X/Fe]_errlogg, [X/Fe]_errvt, [X/Fe]_errsys, [X/Fe]_errtot).

**Supplementary Data 2:** Chemical abundances from individual absorption lines and molecular bands for LMC stars from long-exposure, high-resolution Magellan/MIKE spectra ([link](link)).

The name of the star is followed by the atomic number and ionization of the element measured from the feature. This is followed by the wavelength (in Angstroms), the excitation potential, oscillator strength (log gf), equivalent width, derived abundance, and a flag to indicate whether the measurement is an upper limit. Abundances derived from syntheses of features are indicated in the table, and abundances of the CH molecular band are indicated by the number 106.0 in the second column.